# Inverse Drexhage effect in Epsilon-Near-Zero Substrates

SANDEEP KUMAR CHAMOLI,[1,] MOHAMED ELKABBASH,[2*]

1 Harvard Medical School & Massachusetts General Hospital, Charlestown, MA 02129, USA
2 James C. Wyant College of Optical Sciences, University of Arizona, Tucson, Arizona 85721, USA
*Corresponding author: melkabbash@arizona.edu

**The Drexhage effect, caused by interference between a dipole and its image formed in a substrate, modifies the local density of optical states of quantum emitters which can either enhance or suppress their spontaneous emission rate depending on the dipole orientation and distance from the substrate. Here, we show that for an epsilon-near-zero (ENZ) substrate, the observed orientation and distance dependence of the spontaneous emission rate is reversed compared to metals. This *inverse Drexhage effect* is studied for ideal ENZ and real ENZ substrates compared with ideal and real metallic substrates. ENZ metamaterials consisting of a subwavelength metal-dielectric stack are shown to exhibit the conventional Drexhage effects due to the large optical losses associated with these materials. Our results could find applications in quantum sensing, quantum information, and energy-efficient optoelectronic devices.**

The spontaneous emission (SE) rate is not an inherent property of a given emitter as it strongly depends on the surrounding electromagnetic environment. The emission rate is given by Fermi's golden rule [1]

$$\Gamma_{ij} \propto |M_{ij}|^2 \rho(v_{ij}) \qquad (1)$$

where $\Gamma_{ij}$ is the rate for the transition between the excited state $i$ and lower-energy state $j$; $M_{ij}$, is a matrix element that connects the excited and lower energy levels and is determined by the wavefunctions associated with those levels and $\rho(v_{ij})$ is the density of the optical field at the transition frequency (local density of optical states) [1]. Accordingly, the SE rate of an emitter is proportional to the number of propagating or evanescent electromagnetic modes available to the emitter to radiate into. The Purcell factor ($PF$) quantifies the ratio of the modified radiative decay rate to that in free space $\Gamma_g/\Gamma_0$, where $\Gamma_g$ is the modified SE rate and $\Gamma_0$ is the SE rate in free space for an emitter [2].

The spontaneous emission rate of an emitter separated by a distance $d$ from a mirror periodically oscillates between $PF$ <1 (rate suppression) and $PF$>1 (rate enhancement) as a function of $d$ with the overall modification of the emission rate dropping at large $d$ [3]. This behavior, known as the Drexhage effect, represented one of the earliest quantum electrodynamics experiments. The Drexhage effect arises due to the distance dependent interference between the field radiated by the dipole and its image forming on the mirror substrate. This effect is widely used to probe the local density of optical states and to engineer radiative properties of emitters for applications in quantum optics, light-emitting devices, and cavity QED[4,5]. It provides a simple, geometry-based approach to enhance or suppress emission without the need for nanostructured resonators.

The emission rate strongly depends on the emitter dipole's orientation; A charge $q$ polarizes the surface of its substrate and induces an image charge $q' = -S\, q$ where $S$ is a complex amplitude given by the image coefficient $S = (\varepsilon_{subs} - \varepsilon_0)/(\varepsilon_{subs} + \varepsilon_0)$. For a metallic or a dielectric substrate, the dipole and its image are in an opposite-charge configuration (**Fig. 1a**) [6,7]

Accordingly, for horizontally (vertically) oriented dipoles, the image dipole is out-of-phase (in-phase) and at distances d $\ll \lambda$, this results in $PF < 1$ ($PF > 1$). Note, however, that if the emitter is near the metal substrate ($d < 20\,nm$) the $PF$ increases significantly, regardless of the dipole orientation, due to the excitation of surface plasmon polaritons and lossy surface waves. The high $PF$ here, however, is due to the increase in the non-radiative component and does not improve the quantum efficiency of the emitter defined as:

$$\text{QE} = F_{rad}/(F_{rad} + F_{non-rad}) \qquad (2)$$

In this work, we show that the emission rate from a dipole interacting with an Epsilon Near Zero (ENZ) substrate exhibits the opposite response to a dipole interacting with a metallic substrate. This observation is due to the image coefficient obtained when $Re\,(\varepsilon_{subs}\sim 0)$ leading to $S < 0$. Consequently, the dipole and its image are in a like-charge configuration (**Fig.1b**) [7]. A direct consequence of this "Inverse Drexhage effect" is that for $d \ll \lambda$, a horizontal dipole exhibits an increase in its emission rate and quantum efficiency when it is directly placed on an ENZ substrate. We show that ENZ Metamaterials, consisting of a multilayer stack of metal-dielectric thin-films, exhibit both the Drexhage and Inverse Drexhage effects depending on the dipole orientation [8]. The proposed effect is a promising and lithographically-free approach to control the emission rate and quantum efficiency of quantum emitters with anisotropic dipole orientation such as two-dimensional luminescent materials for quantum sensing of the local density of optical states, quantum information, and energy-efficient optoelectronic devices [9–14].

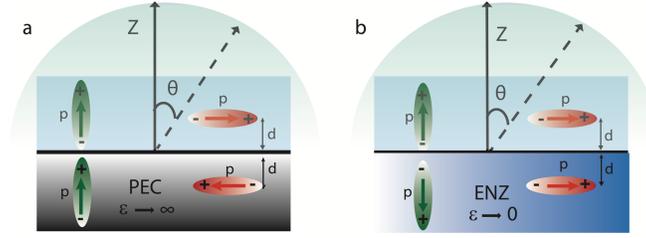

**Figure 1.** A schematic of the system under study. Horizontal and vertical dipole and its image dipole on **(a)** PEC ($\varepsilon \to \infty$) and **(b)** ENZ ($\varepsilon \to 0$) substrate. In the case of PEC ($\varepsilon \to \infty$), the vertical dipole interacts constructively with its image dipole, leading to an enhanced dipole moment, while the horizontal dipole is canceled by its image dipole. In contrast, for ENZ ($\varepsilon \to 0$), the horizontal dipole undergoes constructive interference with its image dipole, resulting in an increased dipole moment, while the vertical dipole is canceled by its image dipole.

**Theoretical model:**

An excited emitter in free space decays to its ground state according to the Larmor formula [1,15]

$$\Gamma_0 = \frac{|\mu|^2 \sqrt{\varepsilon_1} \omega^3}{3hc^3} \quad (3)$$

Where $\mu$ is the electric dipole moment; $\omega$ is emission angular frequency; and $\varepsilon$ is the permittivity of the host material. The $PF$ of a dipole placed a distance $d$ above a substrate with a horizontal ($\parallel$, "P-polarization") and vertical ($\perp$, "S-polarization") orientation is given by [1,15]:

$$PF_\perp = 1 - \eta_0 \left(1 - \frac{3}{2} Re \int_0^\infty dk_x \frac{1}{k_z} \left(\frac{k_x}{\sqrt{\varepsilon_1} k_0}\right)^3 (1 + r_P e^{2ik_z d})\right) \quad (4)$$

$$PF_\parallel = 1 - \eta_0 \left(1 - \frac{3}{4} Re \int_0^\infty dk_x \frac{1}{k_z} \frac{k_x}{\sqrt{\varepsilon_1} k_0} \left[1 + r_S e^{2ik_z d} + \left(\frac{k_z}{\sqrt{\varepsilon_1} k_0}\right)^2 (1 - r_P e^{2ik_z d})\right]\right) \quad (5)$$

Where $\eta_0$ is the internal quantum efficiency of the dipole in free space with real permittivity $\varepsilon_1$, $k_0 = \frac{2\pi}{\lambda_0}$ is the magnitude of the wavevector in vacuum at wavelength $\lambda_0$, $k_z = \sqrt{\varepsilon_1 k_0^2 - k_\parallel^2}$ and $k_\parallel = \sqrt{k_x^2 + k_y^2}$, are the wavevector components along the vertical and horizontal or in-plane directions respectively, and $r^P$, $r^S$ represent the amplitude reflection coefficients of the substrate for a P- (TM) or S- (TE) polarized plane wave and can be calculated using the [16]. The normalized dissipated power spectra are represented by the integrands in equations 4 and 5, respectively.

$$\frac{dP_\perp}{dk_x} = \frac{3}{2} \frac{1}{k_z} \left(\frac{k_x}{\sqrt{\varepsilon_1} k_0}\right)^3 (1 + r_P e^{2ik_z d}) \quad (6)$$

$$\frac{dP_p}{dk_x} = \frac{3}{4} \frac{1}{k_z} \frac{k_x}{\sqrt{\varepsilon_1} k_0} \left\{1 + r_S e^{2ik_z d} + \frac{k_z^2}{\varepsilon_1 k_0^2} [1 - r_P e^{2ik_z d}]\right\} \quad (7)$$

The propagating-wave contribution to the modified spontaneous emission is determined by integrating the dissipated power spectrum in the range of $0 < k_x < k_{cf}$, where $k_{cf} = \sqrt{\varepsilon_1} k_0$ is the cutoff wave-vector. The dielectric medium can only permit emissions with $k_x < k_{cf}$ to escape from it into a corresponding observation or emission angle θ for waves propagating from the near field. On the other hand, wave-vectors $> k_{cf}$, correspond to non-radiative modes that do not couple directly to radiation such as Surface Plasmon Polaritons and Lossy surface waves[1]. Radiative enhancement refers to the ratio between the radiative emission from dipole emitters located on substrates to that from dipole emitters located in free space [17],

$$F_{rad,\perp/\parallel} = Re \frac{\int_0^{k_{cf}} dk_x \frac{dP_{\perp/\parallel}}{dk_x}}{\int_0^{k_{cf}} dk_x \frac{dP_0}{dk_x}} \quad (8)$$

where $P_0$ is the normalized dissipated power for emitters in free space.

**Simulation and Results:**

The $PF$ and $F_{rad}$ are calculated through Finite Difference Time Domain (FDTD) full wave simulations using a commercial software Lumerical®. An electric point dipole was placed adjacent $d$ to the substrate to simulate the interaction of emitters with it. The $F_{rad}$ was calculated by integrating the far-field power density over all emission angles θ (see Figure 1). Finally, the external quantum efficiency ($\eta_{ext}$) was extracted as the ratio of the radiative rate to the total emission rate ($PF$), providing a measure of the system's ability to emit light effectively, i.e, $\eta_{ext} = \frac{F_{rad}}{PF}$.

Figure 2a shows the calculated $PF$ for dipoles placed above a perfect electrical conductor (PEC) substrate, where the conventional Drexhage effect is observed: at $d \ll \lambda$, the emission rate is enhanced for the vertical dipole and suppressed for the horizontal dipole due to the image dipole effect [1]. Figure 2b shows the calculated Purcell factor for dipoles above an ideal, lossless, ENZ substrate, where the inverse Drexhage effect is clearly observed: at $d \ll \lambda$, the emission rate is enhanced for the horizontal dipole and suppressed for the vertical dipole. In addition, the periodic oscillations of the emission rate is observed due to change in phase as a function of $d$ [1]. Since there are no losses, $PF$ should be equal to $F_{rad}$ (see Supplementary Figure 1). The $F_{rad}$ is calculated by integrating the far field power over all angles (see supplementary Figure 2) or wave vectors at a given distance from the PEC and ENZ substrate (see equation 8).

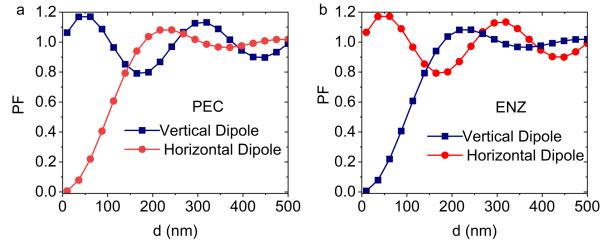

**Figure 2:** $PF$ for vertical and horizontal in case of lossless **(a)** PEC substrate showing the conventional Drexhage effect and **(b)** ENZ substrate showing the inverse Drexhage effect.

We now consider the effect of introducing losses to the ENZ substrate, i.e., $im(\varepsilon) \neq 0$ in Figure 3. For both dipole orientations, introducing losses leads to a sharp increase in the PF (see Figure 3a and b), which, according to the definition in equation 7, does not distinguish between radiative and non-radiative contributions to the emission rate. This increase is significant for $d < 20$ nm which confirms the non-radiative (high-$k$) origin of the high $PF$ where the increase in the decay rate is associated with an increase in the local density of optical states due to the excitation of surface waves, and other dark modes in the substrate. The $F_{rad}$, on the other hand, decreases for horizontal dipoles due to the existence of non-radiative channels for the emitter (see Figure 3d). $F_{rad} > 1$ is still obtained for $im(\varepsilon) < 1$. On the other hand, $F_{rad} \ll 1$ for the vertical dipole at all values for $im(\varepsilon)$ (see Figure 3c). The increase in $F_{rad}$ at higher $im(\varepsilon)$ for the vertical dipole is because the ENZ behavior no longer persists for high $im(\varepsilon)$ and we obtain the conventional Drexhage effect [6,7]. We also derive the dependence of image change $q'$ on the imaginary part of the substrate, showing that for high imaginary permittivity, the sign flips (see supplementary note 1 and equation 1).

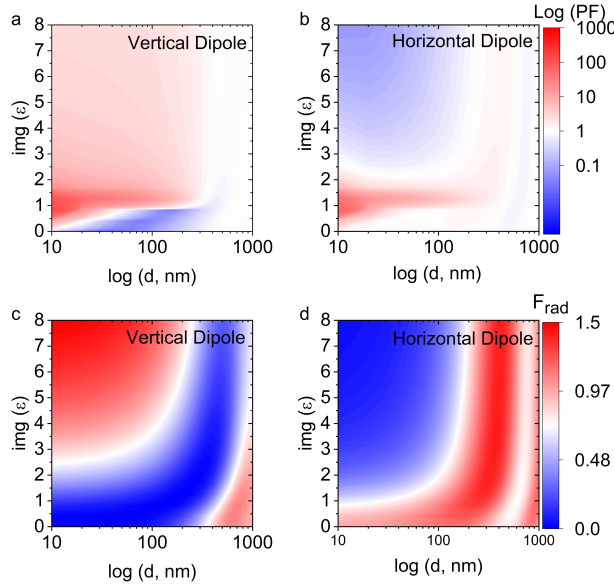

**Figure 3:** $PF$ (logarithmic scale) and $F_{rad}$ for Horizontal dipole and Vertical dipole as a function of dipole distance $d$ (logarithmic scale) and losses in the ENZ substrate keeping $Re(\varepsilon) = 0.01$ at $\lambda = 1324$ nm . **(a)** $PF$ for vertical dipole, **(b)** $PF$ for horizontal dipole, **(c)** $F_{rad}$ for vertical dipole and **(d)** $F_{rad}$ for horizontal dipole.

Next, we compare the $PF$ and $F_{rad}$ as a function of the emitter's distance from the substrate for several substrates—(a) a multilayer (ML) ENZ metamaterial "ENZ-MM (ML)", (b) the same ENZ metamaterial with its permittivity calculated using the effective medium theory "ENZ-MM (EM)" (c) Indium Tin Oxide (ITO) which is a lossy ENZ material at that wavelength and (d) Silver (Ag). The effective permittivity of an ENZ metamaterial consists of a stack of alternating metal (TiN)-dielectric (TiO$_2$) layers shown in Figure 4a is calculated using the effective medium theory [8] (see supplementary note 2). Figure 4b shows the complex permittivity for ITO which is a conductive metal oxide.

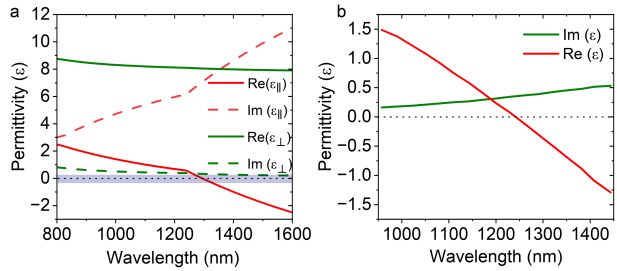

**Figure 4:** Permittivity of **(a)** ENZ MM (TiN (10 nm) +TiO$_2$ (34 nm)) and **(b)** ITO.

Figure 5 examines the *PF* and F$_{rad}$ at 1324 nm for substrates: ITO, ENZ-MM (ML) and Ag. As shown in Figures 5a and 5b, both horizontal and vertical dipoles exhibit a significant *PF* enhancement for ENZ-MM and ITO at short distances with respect to the wavelength. This enhancement arises due to the direct coupling between the dipole field and the surface plasmon modes, ENZ modes, and Ferrel-Berreman modes of ITO [18,19], as well as the high-$\kappa$ bulk plasmon modes available in HMM [8,17]. Ag shows relatively low *PF* due to the low losses at 1324 nm and exhibits a typical Drexhage behavior [17]. Figure 5c shows F$_{rad}$ for horizontal dipoles, where it is evident that lossy ITO substrates exhibit F$_{rad}$ >1 at distances $d \ll \lambda$ around 200 nm, while the vertical dipole (Figure 5d) exhibits a significant drop in F$_{rad}$ demonstrating the inverse Drexhage behavior for horizontal dipoles. We also explored F$_{rad}$ and extracted quantum efficiency $\eta_{ext}$ for the ITO substrate as a function of dipole distance, ranging from near-field to 500 nm, and analyze its behavior within and outside the ENZ wavelength range (see supplementary Figure S3). Conversely the Ag substrate demonstrates the classical Drexhage effect for both the horizontal (Figure 5c) and vertical dipoles (Figure 5d). We also calculate the F$_{rad}$ behavior for Ag and ITO in case of isotropic dipole (see supplementary Figure S4 and S5 respectively). The ENZ-MM is an interesting case. As shown in Fig. 4, these multilayer metamaterials exhibit an anisotropic permittivity with the parallel permittivity crossing epsilon zero region at ~ 1310 nm. The behavior of the horizontal dipole is very similar to the Ag substrate because of the high $im(\varepsilon)$ (Fig. 4a) which makes the multilayer system, for the horizontal dipole, act as a highly lossy ENZ material which exhibits the conventional Drexhage effect (Fig. 3d). On the other hand, the vertical dipole behaves similar to a dipole sitting on a dielectric substrate which also exhibits a conventional Drexhage effect (see Supplementary Figure 6). The behavior of a quantum emitter on the effective medium equivalent of the multilayer system, i.e., ENZ MM (EM), is also modelled to ensure that the results are not due to the multilayer nature of the metamaterial (see Supplementary Figure 7).

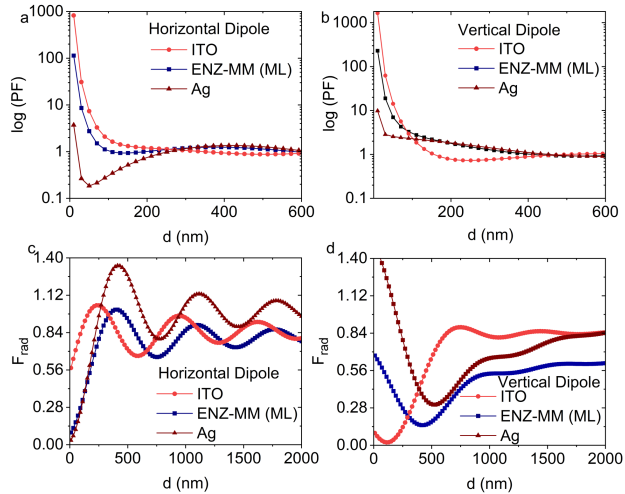

**Figure 5:** *PF* and F$_{rad}$ for horizontal and vertical dipole near ITO, ENZ MM using multilayer system (ML) and Ag substrate at 1324 nm. **(a)** *PF* for Horizontal dipole **(b)** *PF* for Vertical dipole, **(c)** F$_{rad}$ for horizontal dipole and **(d)** F$_{rad}$ for Vertical Dipole.

**Conclusion:**

In conclusion, have demonstrated that ENZ substrates reverse the orientation-dependent behavior of spontaneous emission observed in conventional Drexhage experiments. We showed that horizontal dipoles exhibit enhanced radiative decay rates and external quantum efficiencies when placed near ideal and real ENZ materials, such as ITO, while vertical dipoles are suppressed. In contrast, ENZ metamaterials represent an interesting class as they are highly anisotropic leading to the observation of rich dipole orientation dependence. In particular, the horizontal dipole orientation exhibited a conventional Drexhage behavior, even at ENZ wavelengths, due to the high optical losses of the ENZ metamaterial, while the metamaterial behaved as a dielectric substrate for vertical dipole orientations.

The ability to enhance emission from horizontal dipoles—predominant in 2D materials such as transition metal dichalcogenides [4] using a planar, lithography-free ENZ substrate offers a practical platform for improving the brightness and efficiency of quantum emitters at nanometric distances. This control can also be used for mapping the local density of optical states and for selectively suppressing or enhancing emission in optoelectronic devices, introducing the substrate type as a new degree of freedom for emission engineering. Future work should focus on developing low-loss ENZ metamaterials, where both horizontal and vertical dipoles could exhibit symmetric distance-dependent emission, for example, if the imaginary permittivity of the ENZ metamaterial was below 1 as shown in Fig. 3d. Such behavior would be of major importance for quantum emitters with isotropic orientation (due to symmetry), such as fluorophores embedded in dielectric matrices, by enhancing or suppressing the emission rate of both vertical and horizontal dipole orientations, which can expand the application of the Drexhage effect toward bioimaging, lighting, and quantum photonic integration.

**Funding.** The authors received no specific funding for this work.

**Disclosures.** The authors declare no conflicts of interest.

**Data Availability Statement (DAS).** Data are available upon request from the author.

# Supplementary Information

**Supplementary Figure 1: - Radiative rate ($F_{rad}$) for PEC and ENZ**

Figure 1a presents the radiative rate ($F_{rad}$), showing the same trend as the PF for both dipole orientations since PECs are lossless. Figure 1b presents the radiative rate for the ENZ configuration, following the same trend as the PF for both dipole orientation.

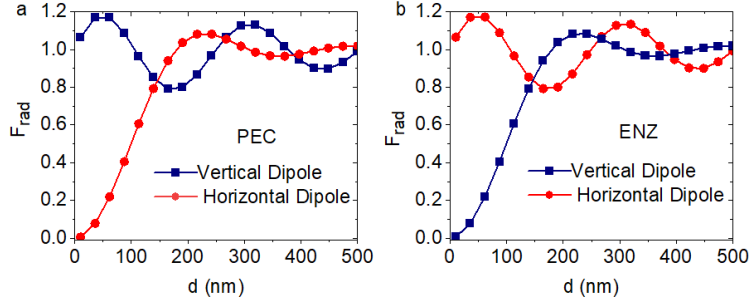

**Figure S1:** (a) $F_{rad}$ for dipoles above a PEC substrate, showing the conventional Drexhage effect, and (b) Frad for dipoles above a ENZ substrate, showing the inverse Drexhage effect

**Supplementary Figure 2: - Far-field power density as a function of emission angle $\theta$**

Figure S3a and Figure S3b show the far-field power density as a function of emission angle $\theta$ and distance for horizontal and vertical dipoles. The results clearly illustrate the angle-dependent interference effect, with the dipole radiative energy varying sinusoidally.

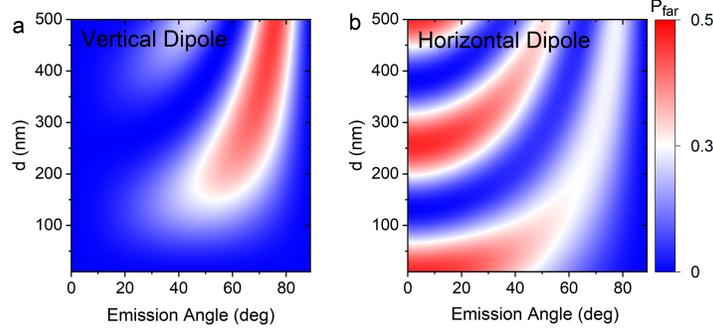

**Figure S2:** Power density per unit solid angle $\theta$, as function of far field emission angle (a) for vertical dipole and (b) horizontal dipole as a function of the collection angle and the inter distance d between the dipole and the substrate.

**Supplementary note 1: - Complex image charge**

We can see the effect of the imaginary part of the substrate explicitly on the image charge modification with the following equation for q':

$$q' = q \frac{(\epsilon_0^2 + Re\epsilon_{subs}^2 + im\ \epsilon_{subs}^2) + i2(\epsilon_0 \epsilon_{subs})}{(\epsilon_0 + im\ \epsilon_{subs})^2 + im\epsilon_{subs}^2} \quad (1)$$

Here, the presence of an imaginary component in substrate permittivity alters both the magnitude and phase of q', leading to a complex image charge that affects field distribution and dipole interactions near the surface. As a result, the dipole image can shift and deform, especially as the dipole moves further from the surface, modifying its orientation and radiation pattern

**Supplementary note 2: - Permittivity calculation for Hyperbolic Metamaterial (HMM)**

The permittivity in Figure 4a for parallel ($\varepsilon_\parallel$) and perpendicular ($\varepsilon_\perp$) permittivity part of HMM is calculated using following expressions [8]:

$$\varepsilon_\parallel = f_m \varepsilon_m + f_d \varepsilon_d \quad (2)$$

$$\varepsilon_\perp = \frac{\varepsilon_m \varepsilon_d}{f_m \varepsilon_m + f_d \varepsilon_d} \quad (3)$$

Where $\varepsilon_d$ and $\varepsilon_m$ are the complex permittivity's of $TiO_2$ and $TiN$, $f_m = \frac{t_m}{t_m + t_d}$ is the fill fraction of $Ag$ and $f_d = 1 - f_m$ is the fill fraction of $TiO_2$, which depends on the thickness of $TiN$ i.e., $t_m$ and thickness of $TiO_2$ i.e., $t_d$. Total thickness of $t_m + t_d$ makes a unit cell.

**Supplementary note 3 and supplementary Figure S3: Extracted quantum efficiency**

For the vertical dipole, Figure S3a shows the $\eta_{ext}$, while Figure S3c presents $F_{rad}$ as the dipole moves from 10 nm to 500 nm across wavelengths from 900 nm to 1400 nm, covering both the ENZ region and pre-ENZ wavelengths of ITO. $F_{rad}$ remains less than 1 almost everywhere across all wavelengths and distances, leading to very low $\eta_{ext}$ for vertical dipole, as expected due to the formation of an image dipole in the ITO that cancels out the vertical dipole's field, suppressing radiative decay and leading low extraction efficiency. On the other hand, for the horizontal dipole, Figure S 3b shows significantly higher $\eta_{ext}$, and $F_{rad}>1$ in Figure S6d due to the inverse Drexhage effect.

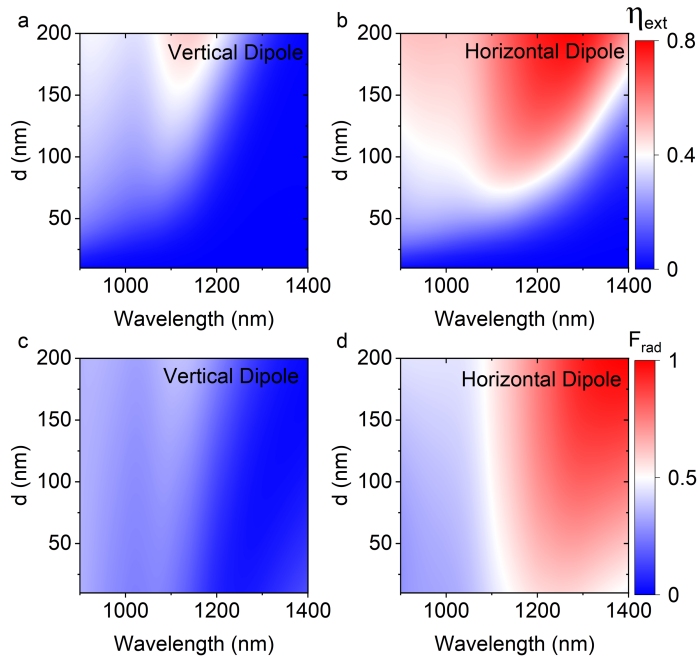

**Figure S3:** $\eta_{ext}$ and $F_{rad}$ on ITO substrate as a function of wavelength and a dipole distance from substrate. (a): $\eta_{ext}$ for vertical dipole, (b): $\eta_{ext}$ for horizontal dipole, (c) $F_{rad}$ for vertical dipole and (d) $F_{rad}$ for horizontal dipole.

**Supplementary Figure 4: - Frad for Isotropic dipole on Ag**

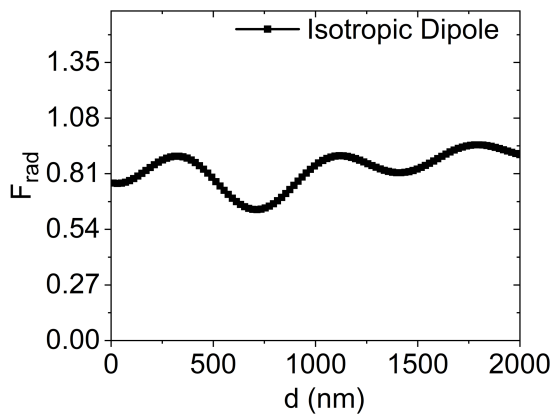

**Figure S4:** $F_{rad}$ as a function of d for isotropic dipole near Ag substrate at a 1324 nm wavelength. The isotropic dipole consistently shows $F_{rad}$ values less than 1 for all dipole distance $d$.

**Supplementary Figure 5: - Frad for Isotropic dipoles on ITO**

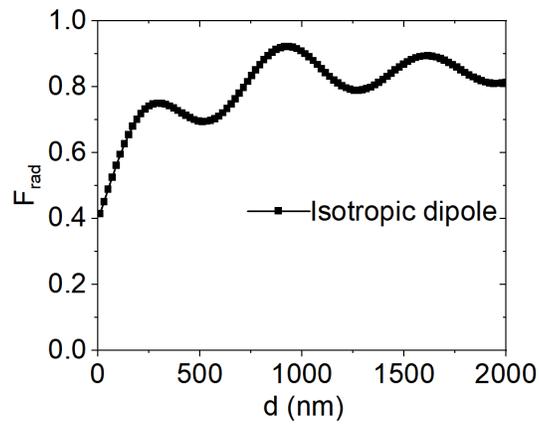

**Figure S5:** $F_{rad}$ as a function of d for isotropic dipole near ITO substrate at a 1324 nm wavelength for isotropic dipoles

**Supplementary Figure 6: - Horizontal dipoles on dielectric with refractive index 4**

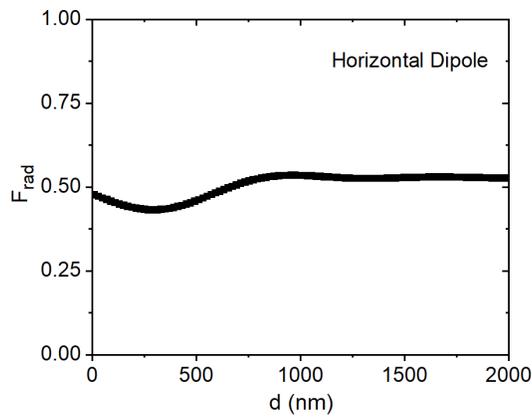

**Figure S6:** $F_{rad}$ as a function of d for isotropic dipole near the dielectric with refractive index 4 at a 1324 nm wavelength for horizontal dipoles

**Supplementary Figure 7: - Comparing PF and $F_{rad}$ using multilayer (ML) system and effective medium (EM) theory for ENZ MM**

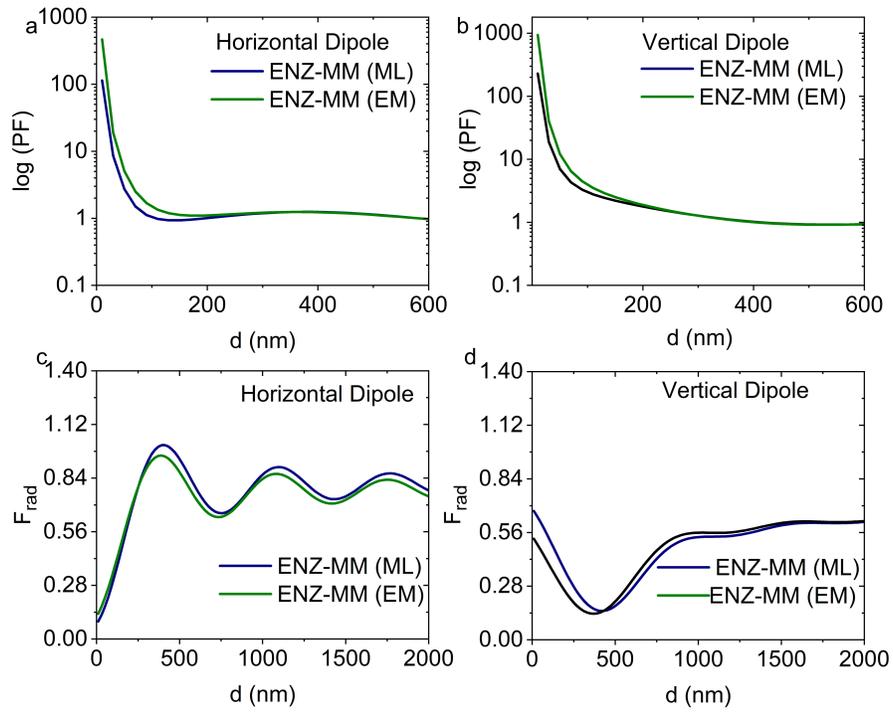

Figure S7: $PF$ and $F_{rad}$ as a function of d for horizontal and vertical dipole near ENZ-MM using ML and EM.